\documentclass[a4paper, prl, twocolumn,showpacs]{revtex4-1}

\pdfoutput=1
\usepackage{amssymb}
\usepackage{amsmath}
\usepackage{epsfig}
\usepackage{color}
\usepackage{graphics, graphicx}
\usepackage{bbold}
\usepackage{psfrag}
\usepackage{mathcomp}
\usepackage{subfigure}
\usepackage{verbatim}

\begin{document}

\date{\today}
\title{Phase separation in a polarized Fermi gas with spin-orbit coupling}

\author{W. Yi and G.-C. Guo}
\address{Key Laboratory of Quantum Information, University of Science and Technology of China,
CAS, Hefei, Anhui, 230026, People's Republic of China}

\begin{abstract}
We study the phase separation of a spin polarized Fermi gas with spin-orbit coupling near a wide Feshbach resonance. As a result of the competition between spin-orbit coupling and population imbalance, the phase diagram for a uniform gas develops a rich structure of phase separated states involving topologically non-trivial gapless superfluid states. We then demonstrate the phase separation induced by an external trapping potential and discuss the optimal parameter region for the experimental observation of the gapless superfluid phases.

\end{abstract}
\maketitle

Spin-orbit coupling (SOC), common in condensed matter systems for electrons, has been considered a key ingredient for many interesting phenomena such as topological insulators \cite{kanereview}, quantum spin Hall effects \cite{xiaoreview}, etc. The recent realization of synthetic gauge field and hence spin-orbit couplings in ultracold atomic systems opens up exciting new routes in the study of these phenomena \cite{gauge1,gauge2}, allowing us to take advantage of the features of the ultracold atoms, e.g. clean environment and highly controllable parameters. In particular, with the Feshbach resonance technique, the effective interaction strength between atoms can be tuned \cite{stoofreview,feshbachexp}. This technique has been applied to study various interesting topics, e.g. the BCS-BEC crossover \cite{crossoverreview}, polarize Fermi gases \cite{polarizedreveiw}, itinerant ferromagnetism \cite{itinerant}, etc.  The introduction of spin-orbit coupling may shed new light on these strongly correlated systems.

Spin-orbit coupled Fermi gas near a Feshbach resonance has recently attracted much theoretical attention \cite{soc1,soc2,soc3,chuanwei,soc4,soc6,iskin}. The SOC has been shown to enhance pairing on the BCS side of the Feshbach resonance \cite{soc3,soc4,soc6}. Furthermore, for a polarized Fermi gas, the SOC introduces competition against population imbalance, which can lead to topologically non-trivial phases \cite{chuanwei,iskin}. Recently, the phase diagrams for a polarized Fermi gas with spin-orbit coupling near a Feshbach resonance have been reported for a uniform gas \cite{iskin}. The phase boundaries have been calculated by solving the gap equation and the number equations self-consistently. However, similar to the case of a polarized Fermi gas near Feshbach resonance \cite{thermo}, due to the competition between different phases, the solutions of the gap equation may correspond to metastable or unstable states. By considering the compressibility criterion \cite{iskin}, the unstable solutions are correctly discarded, while the metastable solutions may survive, rendering the resulting phase boundaries, in particular those representing first order phase transitions, unreliable.

In this paper, we examine in detail the zero temperature phase diagrams for a polarized Fermi gas with Rashba spin-orbit coupling near a wide Feshbach resonance for both the uniform and the trapped cases. To avoid getting metastable or unstable solutions, instead of solving the gap equation, we minimize the thermodynamic potential directly as in Ref. \cite{thermo}. For the uniform gas, we find larger stability regions for the phase separated state at unitarity as compared to the results in Ref. \cite{iskin}. More interestingly, we find that SOC may induce more complicated phase separated states involving gapless superfluid phases that are topologically non-trivial, in addition to the typical phase separated state composed of normal (N) and gapped superfluid (SF) phases. We calculate the stability region for the various phase separated states as well as for the gapless superfluid states, SF state and normal state. We show that there are two distinct gapless phases that differ by the number of crossings their excitation spectra have with the zero energy in momentum space, consistent with previous results \cite{chuanwei,iskin}. These novel gapless phases are stabilized by intermediate SOC strengths; whereas for large enough SOC, the system always becomes a gapped superfluid of `rashbons' \cite{soc3}. We show how these phases can be characterized by their different excitation spectra and momentum space density distributions. We then discuss the phase separation in an external trapping potential, where the various phases naturally phase separate in real space. By examining their respective stability regions, we demonstrate the optimal parameter region to observe the gapless superfluid states in the presence of a trapping potential. For all of our calculations in the paper, we adopt the BCS-type mean field treatment. Although the mean field theory does not give quantitatively accurate results near a wide Feshbach resonance, it is a natural first step for us to qualitatively estimate what phases may be stable, as well as to understand their respective properties. We also note that we have neglected the Fulde-Ferrell-Larkin-Ovchinnikov
(FFLO) phase in our calculations. This is motivated by the fact that the FFLO phase is stable only in a narrow parameter region in the absence of SOC due to competition against other phases \cite{polarizedreveiw}. As SOC introduces new gapless phases into this competition, we do not expect a significant increase in its stability region.

We first consider a uniform three dimensional polarized Fermi gas with Rashba spin-orbit coupling in the plane perpendicular to the quantization axis $z$.  The model Hamiltonian takes the form \cite{soc4,chuanwei,iskin}
\begin{align}
&H-\sum_{\sigma}\mu_{\sigma}N_{\sigma}=\sum_{\mathbf{k},\sigma}\xi_{\mathbf{k}}a^{\dag}_{\mathbf{k},\sigma}a_{\mathbf{k},\sigma}\nonumber\\ &+\frac{h}{2} \sum_{\mathbf{k}}\left(a^{\dag}_{\mathbf{k},\downarrow}a_{\mathbf{k},\downarrow}-a^{\dag}_{\mathbf{k},\uparrow}a_{\mathbf{k},\uparrow}\right) +\frac{U}{\cal V}\sum_{\mathbf{k},\mathbf{k}'}a^{\dag}_{\mathbf{k},\uparrow}a^{\dag}_{-\mathbf{k},\downarrow}a_{-\mathbf{k}',\downarrow}a_{\mathbf{k}',\uparrow}\nonumber\\
&+\sum_{\mathbf{k}}\alpha k_{\perp}\left(e^{-i\varphi_{\mathbf{k}}}a^{\dag}_{\mathbf{k},\uparrow}a_{\mathbf{k},\downarrow} +h.c.\right),
\label{OrgH}
\end{align}
where $\xi_{\mathbf{k}}=\epsilon_{\mathbf{k}}-\mu$, with the kinetic energy $\epsilon_{\mathbf{k}}=\frac{\hbar^2k^2}{2m}$; $\sigma=\{\uparrow,\downarrow\}$ are the atomic spins; $N_{\sigma}$ denotes the total number of particles with spin $\sigma$; $a_{\mathbf{k},\sigma}$($a^{\dag}_{\mathbf{k},\sigma}$) annihilates (creates) a fermion with momentum $\mathbf{k}$ and spin $\sigma$; $\mu_{\sigma}=\mu\pm h/2$ is the chemical potential of the corresponding spin species, and ${\cal V}$ is the quantization volume. The Rashba spin-orbit coupling strength $\alpha$ can be tuned via parameters of the gauge-field generating lasers \cite{gauge2}, while $k_{\perp}=\sqrt{k_x^2+k_y^2}$ and $\varphi_{\mathbf{k}}=\arg{\left(k_x+ik_y\right)}$. In writing Hamiltonian (\ref{OrgH}), we assume s-wave contact interaction between the two fermion species, with the bare interaction rate $U$ renormalized following the standard relation $\frac{1}{U}=\frac{1}{U_p}-\frac{1}{\cal V}\sum_{\mathbf{k}}\frac{1}{2\epsilon_{\mathbf{k}}}$ \cite{crossoverreview}. The physical interaction rate is given as $U_p=\frac{4\pi\hbar^2 a_s}{m}$, where $a_s$ is the s-wave scattering length between the two fermionic spin species.


To diagonalize the Hamiltonian, we make the transformation: $a_{\mathbf{k},\uparrow}=\frac{1}{\sqrt{2}}e^{i\varphi_{\mathbf{k}}}\left(a_{\mathbf{k},+}+a_{\mathbf{k},-}\right)$, $a_{\mathbf{k},\downarrow}=\frac{1}{\sqrt{2}}\left(a_{\mathbf{k},+}-a_{\mathbf{k},-}\right)$, where $a_{\mathbf{k},\pm}$ are the annihilation operators for the dressed spin states with different helicities ($\pm$) \cite{soc3,soc4,chuanwei,soc6,iskin}.  Taking the pairing mean field $\Delta=\frac{U}{\cal V}\sum_{\mathbf{k}}\left\langle a_{-\mathbf{k},\downarrow}a_{\mathbf{k},\uparrow}\right\rangle$ as in the standard BCS-type  theory, we may diagonalize the mean field Hamiltonian in the basis of the dressed spins: $\left\{a_{\mathbf{k},+},a_{-\mathbf{k},+}^{\dag},a_{\mathbf{k},-},a_{-\mathbf{k},-}^{\dag}\right\}^T$. The thermodynamic potential is then evaluated from $\Omega=-\frac{1}{\beta}\ln\text{tr}\left[e^{-\beta(H-\sum_{\sigma}\mu_{\sigma} N_{\sigma})}\right]$, with $\beta=1/k_BT$. In this paper, we will focus on the zero temperature case, for which the thermodynamic potential has the form
\begin{equation}
\Omega=\frac{1}{2}\sum_{\mathbf{k},\lambda=\pm}\left(\xi_{\lambda}-E_{\mathbf{k},\lambda}\right)-{\cal V}\frac{|\Delta|^2}{U},
\label{thermO}
\end{equation}
with the quasi-particle excitation spectrum $E_{\mathbf{k},\pm}=\sqrt{\xi_{\mathbf{k}}^2+\alpha^2k^2_{\perp}+|\Delta|^2+\frac{h^2}{4} \pm 2\sqrt{(\frac{h^2}{4}+\alpha^2k_{\perp}^2)\xi_{\mathbf{k}}^2+\frac{h^2}{4}|\Delta|^2}}$.



\begin{figure}[tb]
\includegraphics[width=8cm]{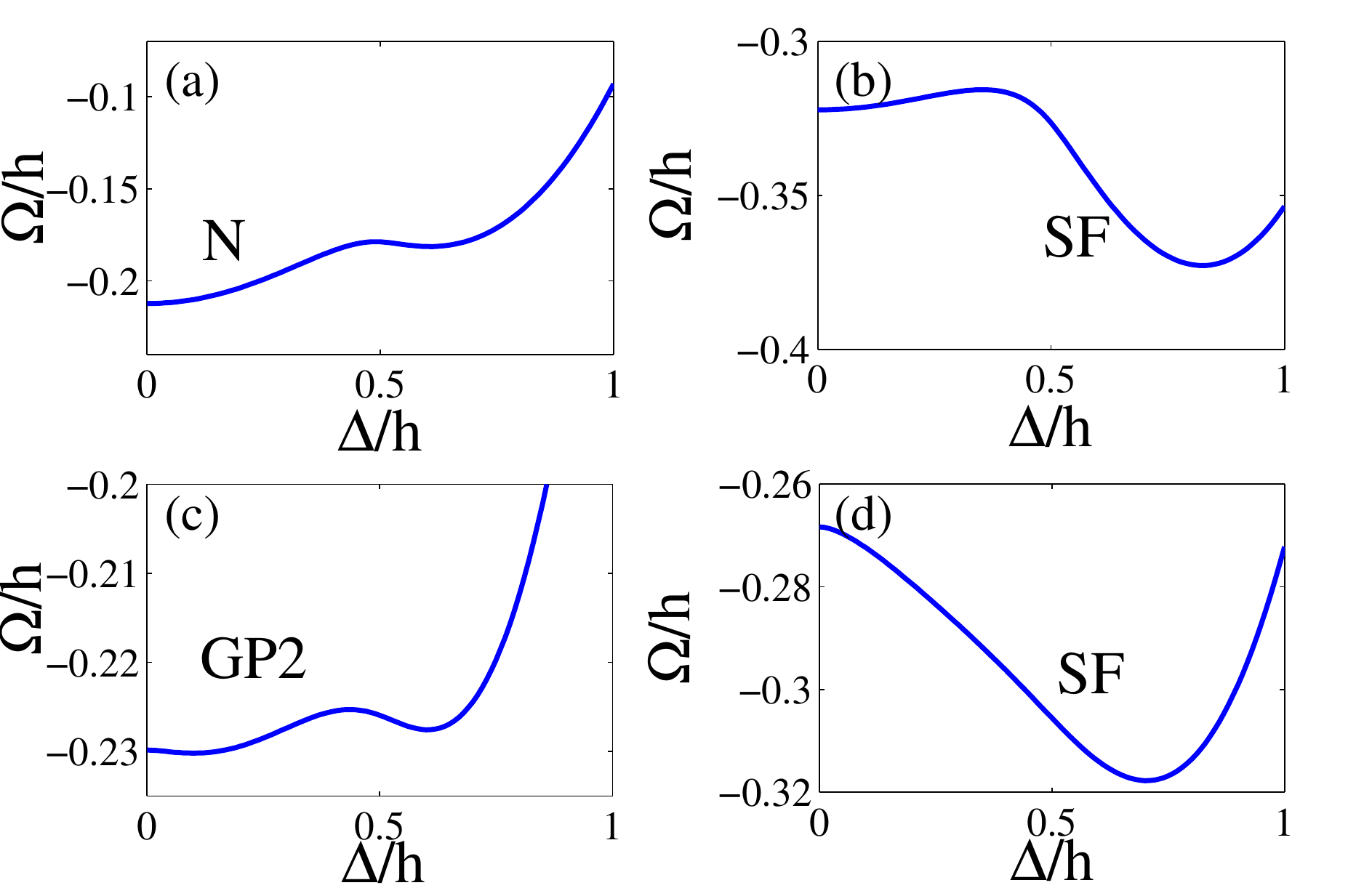}
\caption{Illustration of typical shapes of the thermodynamic potential $\Omega/h$ as a function of order parameter $\Delta/h$ for various phases at unitarity: (a) $\mu/h=0.52$, $\alpha k_h/h=0.1$; (b) $\mu/h=0.7$, $\alpha k_h/h=0.1$; (c) $\mu/h=0.52$, $\alpha k_h/h=0.3$, (d) $\mu/h=0.52$, $\alpha k_F/h=0.6$. The chemical potential $h$ is taken to be the energy unit, while the unit of momentum $k_h$ is defined through $\frac{\hbar^2k_h^2}{2m}=h$. }
\label{omegacomp}
\end{figure}

Before proceeding, let us examine the quasi-particle excitations first and study the conditions for possible gapless phases. We see that at the points in the momentum space where $E_{\mathbf{k},-}$ crosses zero, the quasi-particle excitation becomes gapless while the pairing gap $\Delta$ remains finite. The SOC, together with the population imbalance re-arranges the topology of the Fermi surfaces of the spin species \cite{chuanwei,iskin}. The points of gapless excitations lie on the $k_z$ axis with $k_{\perp}=0$, and are symmetric with respect to the $k_z=0$ plane. More specifically, for $\mu\leq0$, the excitation spectrum has two gapless points $\pm\frac{2m}{\hbar^2}\left(\mu+\sqrt{\frac{h^2}{4}-|\Delta|^2}\right)^{\frac{1}{2}}$, so long as  $\frac{|h|}{2}>\sqrt{\mu^2+|\Delta|^2}$. For $\mu>0$, the excitation spectrum has four gapless points $\left\{\pm\frac{2m}{\hbar^2}\left(\mu+\sqrt{\frac{h^2}{4}-|\Delta|^2}\right)^{\frac{1}{2}},\pm\frac{2m}{\hbar^2}\left(\mu-\sqrt{\frac{h^2}{4}-|\Delta|^2}\right)^{\frac{1}{2}}\right\}$, with $|\Delta|<\frac{|h|}{2}<\sqrt{\mu^2+|\Delta|^2}$; two gapless points $\pm\frac{2m}{\hbar^2}\left(\mu+\sqrt{\frac{h^2}{4}-|\Delta|^2}\right)^{\frac{1}{2}}$, with $\frac{|h|}{2}>\sqrt{\mu^2+|\Delta|^2}$. We identify the superfluid states with two excitation points (GP1) and those with four excitation points (GP2) as different topological phases \cite{chuanwei,iskin}.

\begin{figure}[tb]
\includegraphics[width=8cm]{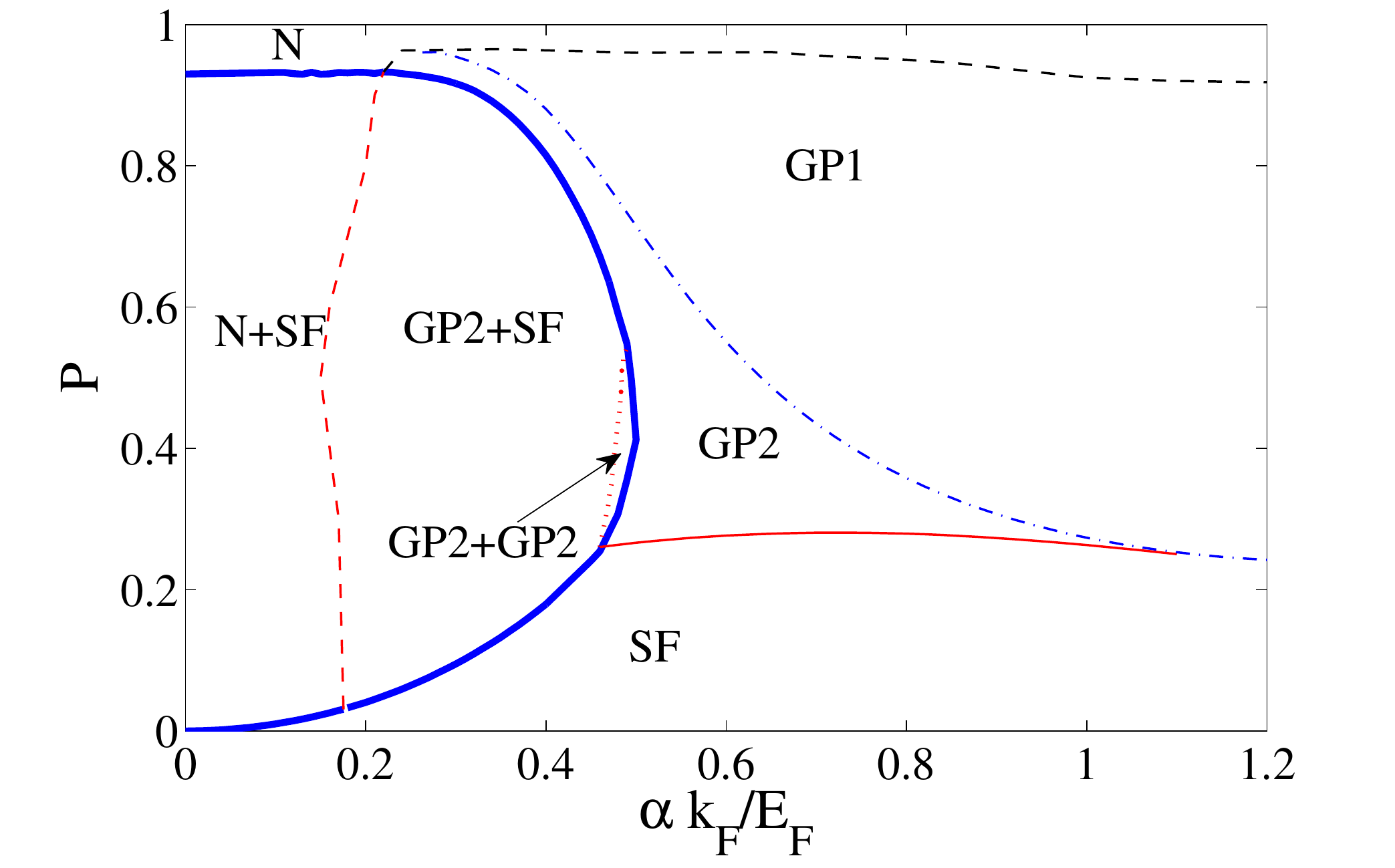}
\caption{Zero temperature phase diagram for a uniform Fermi gas with population imbalance at $(k_Fa_s)^{-1}=0$. Within the bold phase boundaries are the various phase separated states (see text). These phase separated states can be connected with the non-phase separated states by first order phase transitions (solid bold curve). The thin curves represent various second order phase transitions (see text).  Here $k_F=(3\pi^2 n)^{\frac{1}{3}}$, $E_F=\frac{\hbar^2k_F^2}{2m}$, and $n$ is the total density of the system.}
\label{uniformphase}
\end{figure}

We illustrate in Fig. \ref{omegacomp} typical shapes of the thermodynamic potential as a function of $\Delta$ with different parameters. Notably, due to the competition between different phases, a double-well structure appears (see Fig. \ref{omegacomp}(a-c)). Hence the solutions to the gap equation may correspond to the metastable states (local minimum) or the unstable states (local maximum). To make sure that the ground state is achieved, we directly minimize the thermodynamic potential \cite{thermo}.

Another complication comes from the existence of the phase separated state, which must be considered explicitly for a uniform gas. As in the case of  polarized Fermi gases without SOC \cite{polarizedreveiw}, we introduce the mixing coefficient $x$ ($0\leq x\leq 1$),  and the thermodynamic potential becomes
\begin{equation}
\Omega=x\Omega(\Delta_1)+(1-x)\Omega(\Delta_2),
\label{septhermo}
\end{equation}
where $\Delta_i$ ($i=1,2$) is the pairing gap for the $i$th component state. Note that due to SOC, we now have the possibility of a phase separated state of two distinct superfluid states (see Fig. \ref{omegacomp}(c)). The number equations of the phase separated state become
\begin{equation}
N_{\sigma}=x\left.\frac{\partial\Omega}{\partial \mu_{\sigma}}\right|_{\Delta=\Delta_1}+(1-x)\left.\frac{\partial\Omega}{\partial \mu_{\sigma}}\right|_{\Delta=\Delta_2}.
\label{sepnumber}
\end{equation}

Minimizing the thermodynamic potential Eq. (\ref{septhermo}) with respect to $\Delta_i$ and $x$ while implementing the number constraints Eq. (\ref{sepnumber}), we map out the phase diagram for a uniform polarized Fermi gas with SOC at $(k_Fa_s)^{-1}=0$. Fig. \ref{uniformphase} illustrates the resulting phase boundaries in the plane of $(P,\alpha k_F/E_F)$, where the polarization $P=\frac{N_{\uparrow}-N_{\downarrow}}{N_{\uparrow}+N_{\downarrow}}$. When the SOC is off ($\alpha=0$), the system remains in a phase separated state of normal and gapped superfluid (PS1) up to $P\sim 0.93$ before it becomes a normal state via a first order phase transition. This is consistent with previous mean field calculations for a polarized Fermi gas \cite{polarizedreveiw,parish}, while different from the result in Ref. \cite{iskin}. As the SOC strength $\alpha$ increases, a rich structure of different phases shows up, e.g. gapped superfluid phase (SF), gapless superfluid phases with different Fermi surface topology (GP1 and GP2), and notably, various phase separated states. These phase separated states are confined by a phase boundary of first order phase transition (bold curve in Fig. \ref{uniformphase}). In addition to the typical PS1 phase, we now have a phase separated state with  GP2 and SF phases (PS2), and a phase separated state of two distinct GP2 phases (PS3). As $\alpha$ increases, the system can undergo second order phase transitions from PS1 to PS2 and then to PS3 for intermediate $P$ and $\alpha$. Assuming $|\Delta_1|<|\Delta_2|$, the phase boundaries between them can be determined by imposing $\Delta_1=0$ (PS1 and PS2) and $\frac{h}{2}=|\Delta_2|$ (PS2 and PS3), respectively. These phase separated states finally become unstable and give way to single component superfluid phases as $\alpha$ becomes large. The phase boundaries between these single component states are determined by setting $\frac{h}{2}=|\Delta|$ (SF and GP2), $\frac{h}{2}=\sqrt{\mu^2+|\Delta|^2}$ (GP2 and GP1), and $\Delta=0$ (N and GP1), respectively. When $\alpha$ is large enough, the stability region of the GP2 phase decreases and finally vanishes at a tri-critical point ($\mu=0$), beyond which only GP1, SF and normal phase may exist. Note that beyond the tri-critical point, the chemical potential $\mu$ becomes negative, and the phase boundary between GP1 and SF will bend upwards so that in the large $\alpha$ limit the SF phase becomes dominant in the phase diagram.


\begin{figure}[tb]
\includegraphics[width=8cm]{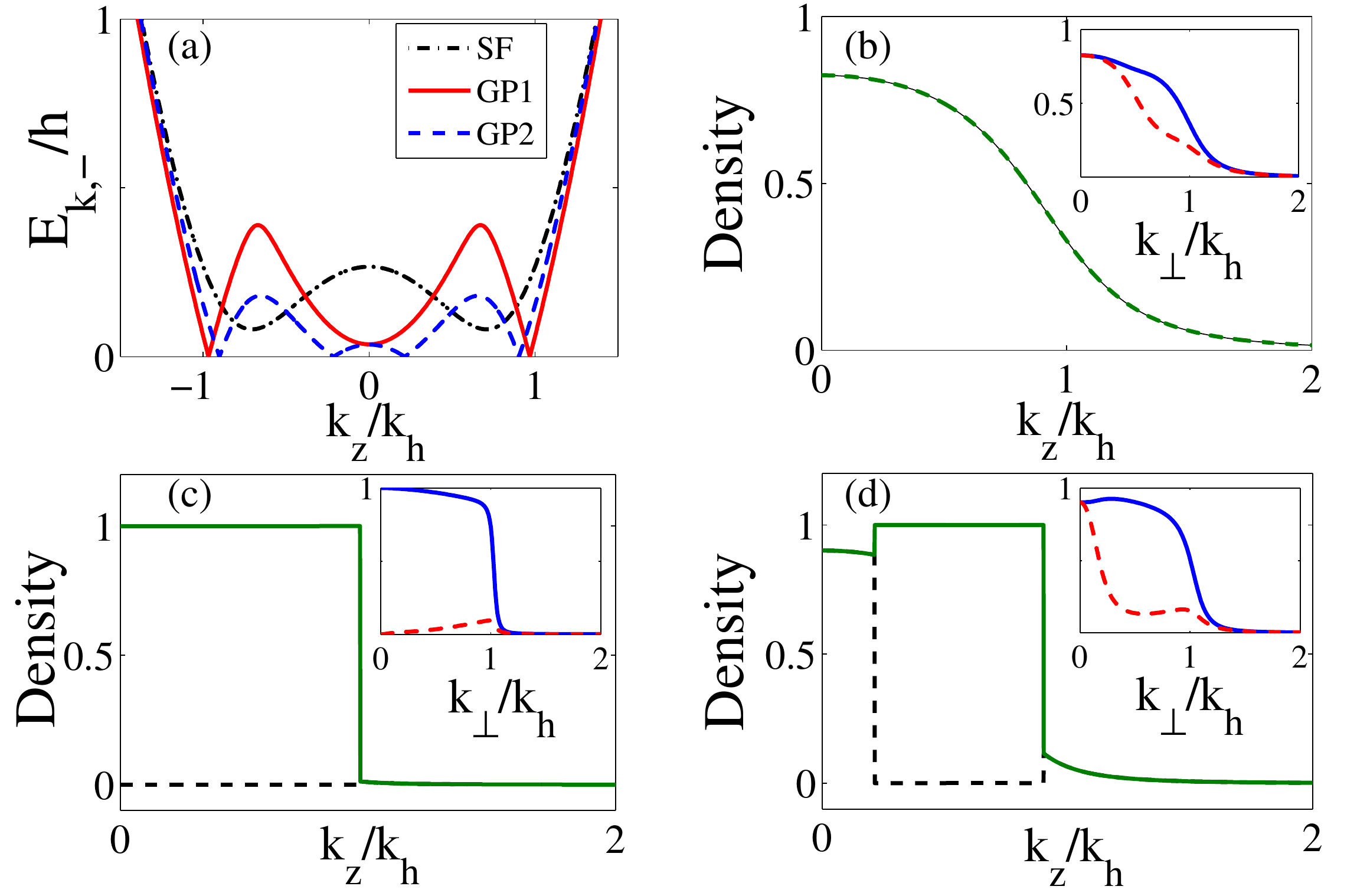}
\caption{Typical excitation spectrum and momentum space density distribution for different phases. (a) Lower branch of the excitation spectra for GP1 (solid), GP2 (dashed) and SF (dash-dotted) phases; (b-d) Density distribution in momentum space for spin-up (solid) and spin-down (dashed) species along $k_{\perp}=0$ and $k_z=0$ (inset), for (b) $\alpha k_h/h=0.35$, $\mu/h=0.5$ (SF); (c) $\alpha k_h/h=0.35$, $\mu/h=0.45$ (GP1); (d) $\alpha k_h/h=0.45$, $\mu/h=0.43$ (GP2), respectively.}
\label{kspace}
\end{figure}

To characterize the properties of the different phases, we calculate the excitation spectrum and number distribution in momentum space for SF, GP1 and GP2 states (see Fig. \ref{kspace}). Several interesting observations are in order. Firstly, the gapless phases leave their signatures in the momentum space density distribution. For $k_{\perp}=0$ and $|k_z|\in \left[\min\left(0,-\sqrt{\mu-\sqrt{\frac{h^2}{4}-|\Delta|^2}}\right),\sqrt{\mu+\sqrt{\frac{h^2}{4}-|\Delta|^2}}\right]$, the minority spin component vanishes, and pairing does not occur in this region. This is reminiscent of the momentum space phase separation of a breached pairing phase in the polarized Fermi gas \cite{bp1}, though now the unpaired region lies only on the $k_z$ axis. Away from $k_z$ axis, the occupation of the minority spin recovers from zero gradually, leaving a signature which may be detected in the time of flight imaging experiment \cite{iskin}. Secondly, for finite $\alpha$, both the gapless and the gapped superfluid phases can support population imbalance, which can be seen from the density distribution along $k_{\perp}$ (see Fig.  \ref{kspace} insets). Indeed as we will see later, for large enough $\alpha$, we may expect no phase separation even in the presence of a harmonic trapping potential. The atoms in the trap will all be in the superfluid phase induced by SOC.

\begin{figure}[tb]
\includegraphics[width=8cm]{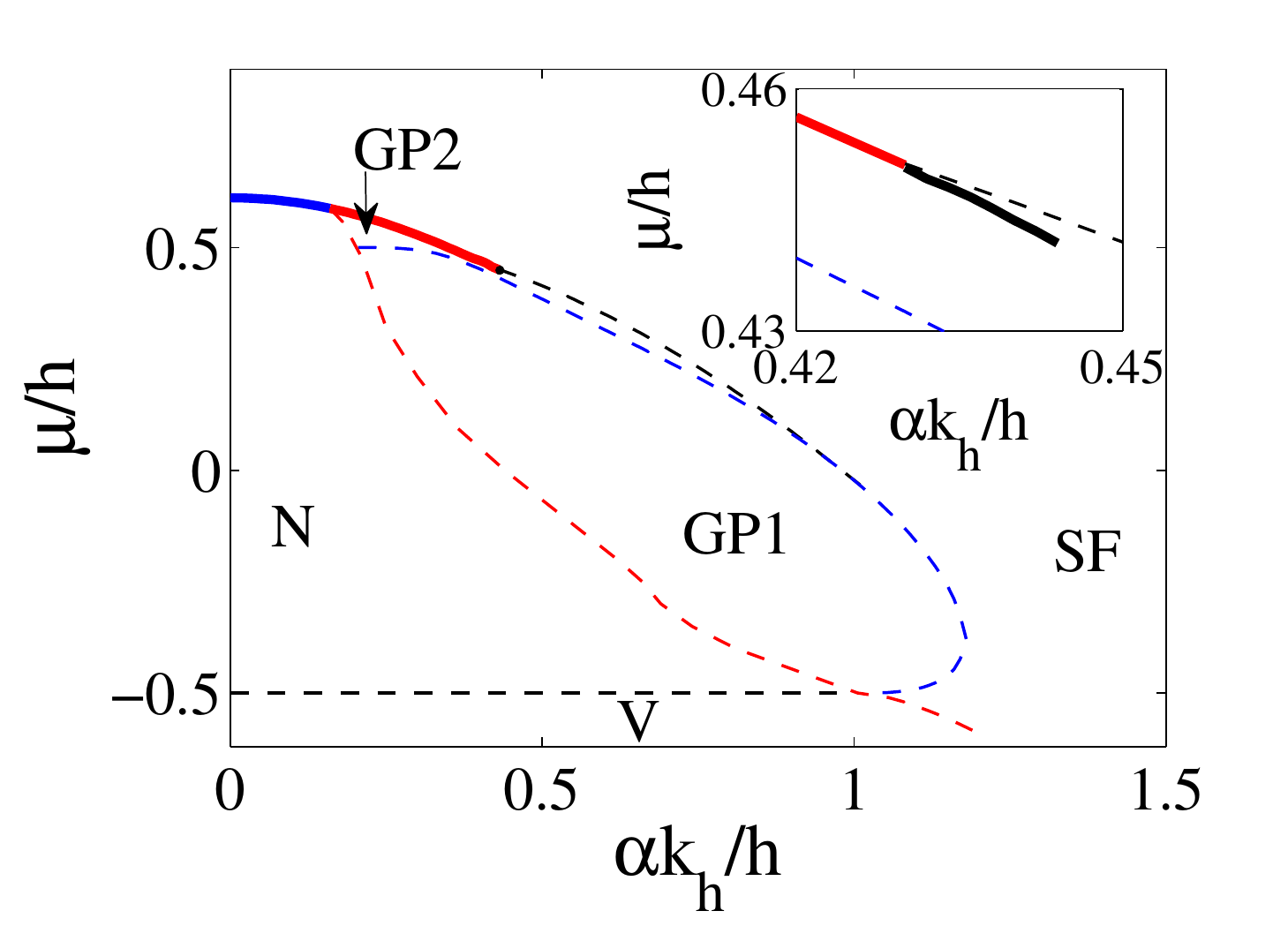}
\caption{Phase diagram in the $(\mu/h,\alpha k_h/h)$ plane at $(k_ha_s)^{-1}=0$. While the second order phase transitions are in dashed thin curves, the first order phase transitions are shown in solid bold curves, which end at the point where the double-well structure in the thermodynamic potential disappears (inset). The boundary for vacuum (V) is determined by setting the chemical potential of the majority spin species to vanish in the normal phase.}
\label{phasetrap}
\end{figure}

To understand the spatial distribution of the various phases in a trapping potential, we calculate the phase diagram as a function of $(\mu/h,\alpha k_h/h)$ at unitarity (Fig. \ref{phasetrap}), where $k_h$ is defined in the caption of Fig. \ref{omegacomp}. Under the Local Density Approximation (LDA) while assuming both spin species experience the same harmonic potential, the local chemical potential $\mu(\mathbf{r})$ can be related to that at the center of the trap $\mu$ as $\mu(\mathbf{r})=\mu-V(\mathbf{r})$, where $V(\mathbf{r})$ gives the trapping potential. Thus a downward vertical line in Fig. \ref{phasetrap} represents a trajectory from a trap center to its edge, with the chemical potential at the trap center fixed by that at the starting point of the line. In Fig. \ref{phasetrap}, consistent with Fig. \ref{uniformphase}, the GP2 phase only exists in a small parameter region in the trap, while there appears to be considerable stability regions for the GP1 phase. When $\alpha$ is small, the Fermi gas in the trap will phase separate into two regions, SF at the core, normal phase (N) towards the edge. At intermediate $\alpha$, the gapless phases GP2 and GP1 may appear either near the center of the trap or as a ring between the SF core and the normal edge, depending on the chemical potentials. Note that the boundary of the first order phase transition between PS3 and GP2 (dotted thin curve in Fig. \ref{uniformphase}) corresponds to a small scale structure here (Fig. \ref{phasetrap} inset), where a first order phase transition (bold black curve) exists between two distinct gapless superfluids, both in the GP2 phase. However, this region is found to be small at unitarity and only increases slightly towards the BCS side of the resonance. It is therefore difficult to observe this phase transition in the trapped Fermi gas in the parameter region that we considered. For large SOC beyond the tip of the GP1-SF phase boundary, there is only the SF phase in the phase diagram, and hence we will have only the SOC induced SF phase in the trap for large enough SOC.


\begin{figure}[tb]
\includegraphics[width=9cm]{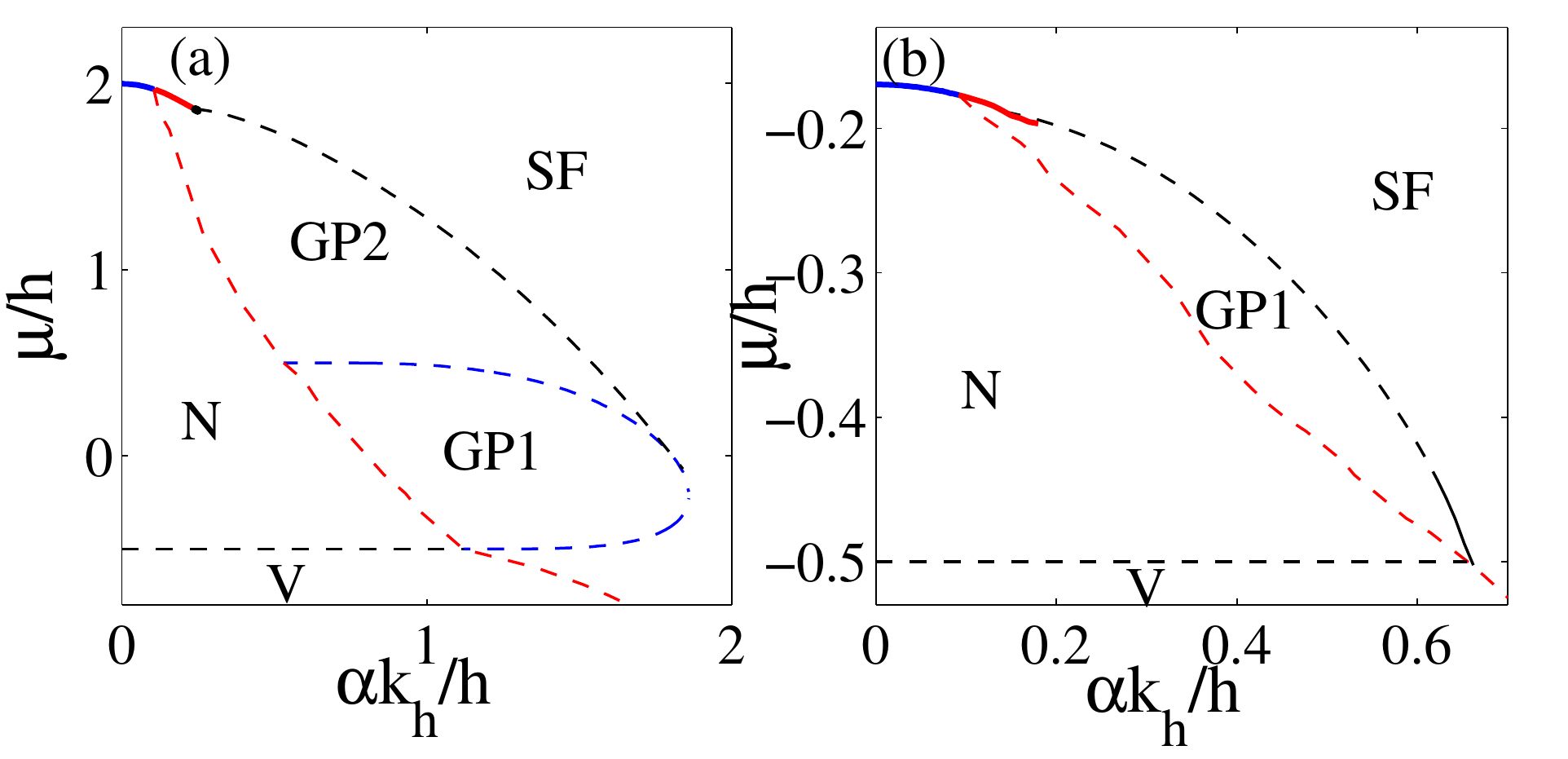}
\caption{Phase diagram in the $(\mu/h,\alpha k_h/h)$ plane at (a) $(k_ha_s)^{-1}=-1$ and (b) $(k_ha_s)^{-1}=0.5$. First order phase transitions are shown in solid bold curves, while second order phase transitions are in dashed thin curves.}
\label{phasetrapmore}
\end{figure}

We have also calculated the phase diagram in $(\mu/h,\alpha k_h/h)$ plane away from the resonance point. On the BCS side (Fig. \ref{phasetrapmore}(a)), the stability region for the GP2 phase increases considerably. It is therefore desirable to prepare the system on the BCS side of the resonance for the observation of GP2 phases. On the BEC side (Fig. \ref{phasetrapmore}(b)), the GP2 phase vanishes from the phase diagram altogether for $\mu<0$, consistent with our previous discussion.

In summary, we have calculated in detail the phase diagrams near a wide Feshbach resonance for a polarized Fermi gas with Rashba spin-orbit coupling. We find that the competition among pairing, polarization and SOC gives rise to a rich structure of phases and phase separations involving topologically non-trivial phases. From the phase diagrams for both uniform and trapped systems, we find that the interesting gapless superfluid phases are most likely to be observed in an experiment with moderate polarization and SOC strength.


We would like to thank L.-M. Duan for helpful discussions. This work was supported by NFRP 2011CB921200 and 2011CBA00200,
NNSF 60921091, and The Fundamental Research Funds for the Central Universities WK2470000001.

\end{document}